\documentclass[aps,prl,twocolumn,superscriptaddress,superscriptreference]{revtex4-2}
\usepackage{amsmath,bbold,bm,amssymb,scalerel,mathtools}
\usepackage{graphicx}
\usepackage{color}
\usepackage{enumitem}
\usepackage{algorithm,algpseudocode}
\usepackage{multirow}
\usepackage{colortbl,booktabs}
\usepackage{placeins}
\usepackage[usenames,dvipsnames]{xcolor}
\usepackage{appendix}
\usepackage{tikz}
\usepackage[colorlinks, linkcolor=blue!50!black, urlcolor=blue!50!black, citecolor=blue!50!black]{hyperref}
\tikzset{%
	every neuron/.style={
		circle,
		draw,
		minimum size=0.5cm
	},
	neuron missing/.style={
		draw=none, 
		scale=1.5,
		text height=0.333cm,
		execute at begin node=\color{black}$\vdots$
	},
}

\newcommand\equalhat{%
	\let\savearraystretch\arraystretch
	\renewcommand\arraystretch{0.3}
	\begin{array}{c}
		\stretchto{
			\scalerel*[\widthof{=}]{\wedge}
			{\rule{1ex}{3ex}}%
		}{0.5ex}\\ 
		=%
	\end{array}
	\let\arraystretch\savearraystretch
}

\newcommand{\gj}[1]{\textcolor{black}{#1}}

\usepackage[normalem]{ulem}

\newcommand\state{\mathcal{S}}
\newcommand\mem{\mathcal{M}}
\newcommand\reward{\mathcal{R}}
\newcommand\policy{\mathcal{P}}

\newcommand\fzero{f_0}

\newcommand\lamLt{\tilde{\lambda}_T}

\begin{document}
	
	\setlength{\belowdisplayskip}{3pt} \setlength{\belowdisplayshortskip}{3pt}
	\setlength{\abovedisplayskip}{3pt} \setlength{\abovedisplayshortskip}{3pt}
	
	\title{Kinetic theory of decentralized learning for smart active matter}
	
	\author{Gerhard Jung}
	
	\affiliation{Universit\'e Grenoble Alpes, CNRS, LIPhy, 38000 Grenoble, France}
	
	\author{Misaki Ozawa}
	
	\affiliation{Universit\'e Grenoble Alpes, CNRS, LIPhy, 38000 Grenoble, France}
	
	\author{Eric Bertin}
	
	\affiliation{Universit\'e Grenoble Alpes, CNRS, LIPhy, 38000 Grenoble, France}

	\date{\today}
	
	\begin{abstract}
		Smart active matter has the ability to control its motion guided by individual policies to achieve collective goals. We introduce a theoretical framework to study a decentralized learning process in which agents can locally exchange policies to \gj{adapt their behavior and maximize a predefined reward function}. We use our formalism to derive explicit hydrodynamic equations for the policy dynamics. We apply the theory to two different microscopic models where policies correspond either to fixed parameters similar to evolutionary dynamics, or to state-dependent controllers known from the field of robotics. We find good agreement between theoretical predictions and agent-based simulations. By deriving fundamental control parameters and uncertainty relations, our work lays the foundations for a statistical physics analysis of decentralized learning.
	\end{abstract}
	
	\maketitle
	
	Decentralized optimization processes have been intensively studied in interdisciplinary fields using methods inspired from statistical physics, including the study of evolutionary and population dynamics in biology \cite{drossel2001biological,sella2005application,houchmandzadeh2011fixation,chia2011statistical} and the investigation of social processes \cite{castellano2009statistical}, such as opinion dynamics \cite{lorenz2007continuous}. 
	In addition, decentralized adaptation raises a growing interest in robotics and engineering \cite{watson2002embodied,bredeche2018embodied,long2018towards,doi:10.1177/1059712320930418,ben2023morphological,bredeche2022social}, in particular to find more stable policies than typical centralized protocols \cite{long2018towards,ben2023morphological}.
	Robustness and flexibility may be enhanced by drawing inspiration from collective animal behavior \cite{verdoucq2022bioinspired}.
	\gj{Standard experiments on robotic swarms typically involve around a hundred robots, and their numbers are increasing. This growth calls for a statistical physics approach to describe assemblies of small robots interacting through information exchange~\cite{ben2023morphological}.}
	
	The idea of experimentally realizing large swarms of microrobots \cite{wang2024robo,ma2024smarticle,ozkan2021collective} as a way to build responsive or programmable metamaterials
	\cite{Li2021programming,zhou2022programmable,kotikian2019untethered,zeravzic2017colloquium} is also at the core of the emerging field of smart active matter \cite{PhysRevE.97.042604,pishvar2020foundations,cichos2020machine,kaspar2021rise,levine2023physics,goldman2024robot,nasiri2024smart}.
	Along this soft robotics perspective, one tries to build large numbers of extremely simplified microrobots
	\cite{whitesides2018soft,majidi2019soft,tsang2020roads,xu2022locomotion}, whose size ranges from the granular scale \cite{saintyves2024self,savoie2019robot} down to the colloidal one \cite{Liu2023colloidal,zeravzic2017colloquium}, and whose main features result from the physical properties of the soft material they are made of
	\cite{stern2020supervised,stern2020continual,dillavou2022demonstration,mandal2024learning}.
	These soft microrobots may integrate simple computation capabilities \cite{garrad2019soft}, or
	may be controlled by an external computer through machine learning using, e.g., reinforcement learning techniques
	\cite{muinos2021reinforcement,cichos2023artificial,colabrese2017flow,gustavsson2017finding,durve2020learning,falk2021learning,sankaewtong2023learning,zou2024adaptative,xiong2024enabling,grauer2024optimizing}.
	Although their starting points differ, both hard and soft robotics aim at building large assemblies of autonomous microrobots with communication and computation capabilities
	leading to adaptative collective behaviors \cite{ben2023morphological,bredeche2022social,kaspar2021rise,levine2023physics,majidi2019soft,mo2023challenges,cazenille2024signaling}.
	This goal is also shared by the emerging topic of biological metamaterials, consisting for instance of insects aggregates  \cite{tennenbaum2016mechanics,wagner2022computational}. 
	
	The ongoing experimental and numerical development of smart active matter necessitates appropriate theoretical approaches to describe its collective properties.
	Early contributions in this direction include models of individual \cite{liebchen2019optimal,piro2021optimal,piro2022optimal,piro2022efficiency,nasiri2023optimal} or
	collective navigation of active particles \cite{borra2021optimal,yang2022autonomous}, or of pedestrians in a dense crowd \cite{echevarria2023body,bonnemain2023pedestrians}.
	Stochastic thermodynamics may also be relevant to describe smart active matter \cite{vansaders2023informational,cocconi2024dissipation}.
	Yet, apart from some first steps \cite{ziepke2022multi,vansaders2023informational}, a theoretical framework to describe the collective behavior of adaptive agents exchanging information with their neighbors \cite{bredeche2022social,ben2023morphological} is still lacking.
	
	In this Letter, we develop a kinetic theory framework to describe decentralized learning in smart active matter with local information exchange. 
	In contrast to usual active matter 
	\cite{ramaswamy2010mechanics,marchetti2013hydrodynamics,ramaswamy2019active}, smart active matter consists of agents capable of responding to external stimuli, communicating  and \gj{adapting their behavior. Through this adaptability, they can learn to solve specific tasks or to attain targeted collective properties.} The behavior of individual agents is determined by a `policy', i.e., a set of rules controlling their dynamics, whose parameters can be tuned.
	Among the modeling approaches used to characterize active matter, kinetic theories have played an important role
	by providing a framework to derive macroscopic hydrodynamic equations from microscopic collision rules \cite{bertin2006boltzmann,bertin2009hydrodynamic,ihle2011kinetic,bertin2013mesoscopic,ihle2014towards}. 
	We describe decentralized learning via the exchange of policies between neighboring agents, akin to collision rules, setting the stage for a kinetic theory framework,
	within which we systematically derive macroscopic hydrodynamic equations for the time-evolution of policies starting from agent-based models.

	\begin{figure}[b]
		\includegraphics[scale=0.6]{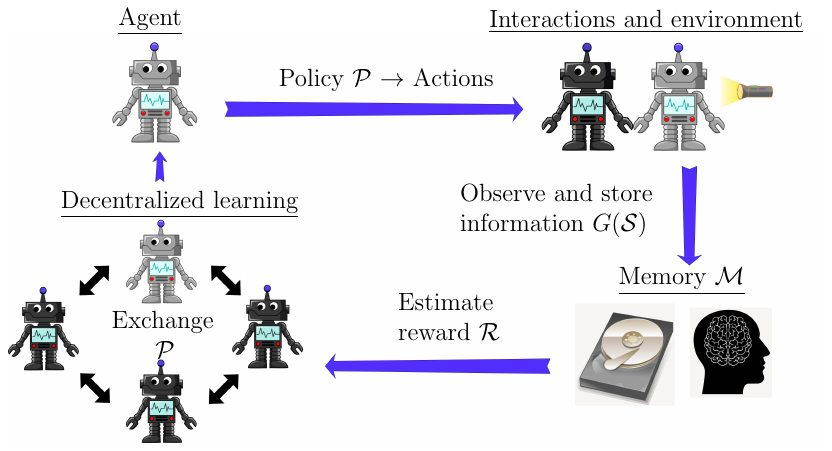}
		\caption{Sketch of the decentralized learning procedure described in the manuscript. }
		\label{fig:sketch}
	\end{figure}

	\emph{Agent-based model of decentralized learning} -- We consider a generic system of $N$ agents in $d$ \gj{spatial} dimensions representing, e.g., small robots \cite{ben2023morphological}.
	Agents are located at position $\bm{x}$, move at velocity $\bm{v}$ with an overdamped dynamics, and diffuse with a positional diffusion coefficient $D_0$.
	Agents sense their environment (e.g., external fields or properties of neighboring particles) to control their dynamical rules.  The set of adjustable parameters controlling the dynamics, \gj{whose values} are specific to each agent, is here generically called the agent's policy $\policy$.
	In the examples below, the policy either corresponds to the agent's angular diffusion coefficient, or to the response of the self-propulsion speed to an external stimulus.
	The state $\state$ of an agent gathers internal physical variables (e.g., orientation, self-propulsion velocity, etc.) and sensors outputs such as measured light intensity.
	Agents can then analyze observed states and extract the \gj{processed information} $G(\state)$, which fluctuates on a characteristic time scale $\tau_G$. The \gj{processed information} is then \gj{filtered and} stored in the `memory' $\mem$  via \gj{the running average on the time scale $\tau_\mem$},
	\begin{equation} \label{eq:mem:dyn}
	\frac{\text{d}\mem(t)}{\text{d}t} = -\frac{1}{\tau_\mem}\, \big(\mem(t) - G(\state(t))\big),
	\end{equation}
	\gj{where $t$ is time. The purpose of the memory is to extract the relevant signal from potentially noisy observations $G(\state)$. This ensures that reward is not awarded for attaining one specific state but rather to promote long-time behavior \cite{ben2023morphological}. Eq.~(\ref{eq:mem:dyn}) is equivalent to a convolution of the processed information $G(\state)$ by an exponential memory kernel of characteristic time scale $\tau_\mem$.  } The quantities $\state$, $\mem$, and $\policy$ may be multi-component vectors. To lighten the presentation of the formalism, we stick to the scalar case, but the multi-dimensional formalism
	is presented in the supplemental material (SM) \cite{suppmat}.
	
	\gj{To enable learning, agents can exchange information about the efficiency of their individual policies. This efficiency is quantified by a {scalar} reward function $\reward(\mem),$ which is evaluated solely based on each agent's individual memory $\mem$. Agents will thus optimize their personal reward, but the reward function {may} be chosen to promote {a specific} target collective behavior (see examples below).}
	Comparing its own reward with other agents enables each agent to learn improved policies (Fig.~\ref{fig:sketch}).
	Agents are therefore randomly selected with rate $\lambda_T=\tau_T^{-1}$, called the teaching rate ($\tau_T$ is the teaching time).
	Once selected, an agent $i$ iterates over all neighbors $n$, chosen in random order within the interaction radius $r_c$.
	For each selected neighbor $n$, agent $i$ becomes a teacher (and agent $n$ the student) with a probability $p_T(\reward_i,\reward_n)$ given by
	\begin{equation} \label{eq:def:PT}
	p_T(\reward_i,\reward_n) = \frac{1}{2}\left( 1+ \tanh\big[\alpha_T \big(\reward_i - \reward_n \big)\big] \right),
	\end{equation}
	where $\reward_i=\reward(\mem_i)$ and $\reward_n=\reward(\mem_n)$.
	Otherwise, agent $n$ is the teacher and agent $i$ is the student.
	In either case, the student updates its policy $\policy_S$ and memory $\mem_S$ by adapting the corresponding values of the teacher, i.e.,
	\gj{$\policy^{\text{new}}_S = \policy_T$ and $\mem^{\text{new}}_S = \mem_T$}.
	The memory of the student is updated together with the policy to make its reward consistent with the updated policy \cite{ben2023morphological}.
	The parameter $\alpha_T$ in Eq.~(\ref{eq:def:PT}) controls the sharpness of $p_T$, and may be interpreted as the precision with which agents evaluate and communicate the reward $\reward$. \gj{In Eq.~(\ref{eq:def:PT}) only reward differences matter and agents are lacking a concept of an absolute reward, which highlights the decentralized nature of the optimization.} 
	To enhance adaptability, we also include a small rate of spontaneous policy changes. By analogy with evolutionary processes in biology, we call these spontaneous changes `mutations' \cite{chardes2023_mutations} and model them as a diffusion process in policy space with diffusion coefficient $D_{\rm mut}$.
	
	\emph{Kinetic theory of decentralized learning} --
	To obtain a statistical description of the above agent-based model and derive the time-evolution for the statistics of policies, we introduce a kinetic theory framework. \gj{This approach is natural since the learning events correspond to binary, intermittent interactions, which can be treated in} analogy with binary collisions in usual kinetic theories.
	In this framework, an assembly of a large number $N$ of agents is described by a single-agent phase-space density $f(\state,\mem,\policy,\bm{x},t)$ characterizing the probability to find an agent at time $t$ at position $\bm{x}$, in state $\state$, with memory $\mem$ and policy $\policy$. 
	The time-evolution formally reads 
	\begin{equation}\label{eq:KT}
	\frac{\partial f}{\partial t} = \mathcal{I}_{\rm phys}[f] + \mathcal{I}_{\rm mem}[f] + \gj{\mathcal{I}_{\rm learn}[f]}.
	\end{equation}
	\gj{The term $\mathcal{I}_{\rm phys}[f]$ describes the physical dynamics of the state $\state$ and the position $\bm{x}$, e.g., advection and diffusion, angular dynamics and physical interactions.} 
	$\mathcal{I}_{\rm mem}[f]$ in Eq.~(\ref{eq:KT}) encodes the memory dynamics given in Eq.~(\ref{eq:mem:dyn}).
	Finally, the term $\mathcal{I}_{\rm learn}[f]$ accounts for the dynamics of the policy $\policy$, which includes diffusive mutations and communication-driven decentralized learning,
	\begin{equation} \label{eq:KT:policy}
	\gj{\mathcal{I}_{\rm learn}[f]} = D_{\rm mut} \frac{\partial^2 f}{\partial \policy^2} + \gj{\mathcal{I}_{\rm teach}[f]}.
	\end{equation}
	\gj{The teaching term $\mathcal{I}_{\rm teach}[f]$ describes the binary teaching events introduced in the agent-based model.
		Mathematically, the term $\mathcal{I}_{\rm teach}[f]$ is a bilinear integral of the phase-space density $f$.}
	The explicit, but lengthy, expression of $\mathcal{I}_{\rm teach}[f]$ is reported in the End Matter (EM) section.
	We assume in this manuscript a time scale separation $\tau_G \ll \tau_\mem \ll \tau_T$, i.e., a hierarchy of time scales separating \gj{processed information dynamics} $\tau_G$, memory $\tau_\mem$ and teaching events $\tau_T$. \gj{In other words, agents have a memory time $\tau_\mem$ long enough to filter the signal from noisy observations and are learning sufficiently slow to enable the memory to converge towards its steady-state value for a fixed policy. } Information advection by particle motion therefore dominates over information propagation due to the finite communication range between agents.
	Within this approximation, our theory is thus purely local. However, it can be generalized to consider non-local communication or overlapping time scales.
	
	To characterize the policy statistics, we assume that \gj{$f$ is approximately Gaussian in $\mem$ and $\policy$}, \gj{but this assumption can also be relaxed (see EM)}. Consequently, the reduced phase-space density
	\begin{equation}\label{eq:phi0}
	\varphi_0(\policy,\bm{x},t) = \int d\state \int d\mem f(\state,\mem,\policy,\bm{x},t)
	\end{equation}
	can be described by its lowest-order moments. 
	We also define the policy-dependent average memory,
	\begin{equation}\label{eq:mem}
	\mu_\mem(\policy,\bm{x},t) = \varphi_0^{-1} \! \int \!d\state \!\int \!d\mem \mem f(\state,\mem,\policy,\bm{x},t).
	\end{equation}
	The reduction steps from $f(\state,\mem,\policy,\bm{x},t)$ to $\varphi_0(\policy,\bm{x},t)$ and $\mu_\mem(\policy,\bm{x},t)$ are described in the EM section.
	\gj{To derive closed hydrodynamic equations for the moments introduced above}, we expand $\mu_\mem(\policy,\bm{x},t)$ around a predefined policy value $\policy_*$. \gj{To ensure that this approximation is consistent, $\policy_*$ should be a rough estimate of the optimal policy}, yielding
	\begin{equation} \label{eq:mu:mem}
	\mu_\mem(\policy,\bm{x},t)=\mu_{\mem}^{(0)}(\bm{x},t)+\mu_{\mem}^{(1)}(\bm{x},t) (\policy-\policy_*)+\dots
	\end{equation}
	The quantities $\mu_{\mem}^{(n)}(\bm{x},t)$ are model-specific, and can be determined explicitly under the time scale separation hypothesis \cite{suppmat}.
	We further introduce the agent density $\rho(\bm{x},t)=\int d\policy \varphi_0(\policy,\bm{x},t)$, the average policy $\mu(\bm{x},t)$ and its variance, called {\it diversity}, $\sigma^2(\bm{x},t)$, defined as
	\begin{align}\label{eq:mu}
	\mu(\bm{x},t) &= \rho^{-1} \int d\policy \policy \varphi_0(\policy,\bm{x},t),\\\label{eq:var}
	\sigma^2(\bm{x},t) &= \rho^{-1}\int d\policy (\policy - \mu(\bm{x},t)) ^2 \varphi_0(\policy,\bm{x},t).
	\end{align}
	For spatially uniform fields, $\rho(t)$, $\mu(t)$, and $\sigma^2(t)$, the evolution equations take the form, 
	\begin{align} \label{eq:evol:mu}
	\frac{d \mu}{d t} &= \rho  \sigma^2 F_1\left(\mu,\sigma^2;\mu_{\mem}^{(n)}\right),\\
	\label{eq:evol:sigma}
	\frac{d \sigma^2}{d t} &= 2 D_{\rm mut} - \rho \sigma^4 F_2\left(\mu,\sigma^2;\mu_{\mem}^{(n)}\right),
	\end{align}
	where $F_m\big(\mu,\sigma^2;\mu_{\mem}^{(n)}\big)$ with $m=1,2$ are model-specific polynomials \cite{suppmat} depending on the parameters $\mu_{\mem}^{(n)}(t)$ introduced in Eq.~(\ref{eq:mu:mem}), which can be determined independently.
	The closed Eqs.~(\ref{eq:evol:mu}) and (\ref{eq:evol:sigma}) for the policy dynamics are one of the major results of this work.
	The full hydrodynamic equations including space-dependent fields are derived for two specific models in the SM \cite{suppmat}.
	\gj{An explicit model where space-dependence is taken into account is given below.}
	Importantly, we can already draw some non-trivial general conclusions from Eqs.~(\ref{eq:evol:mu}) and (\ref{eq:evol:sigma}): (i) The \gj{time} derivative $\dot{\mu}(t)$ is proportional to the density $\rho(t)$, highlighting the many-body character of policy dynamics. Similarly, it is proportional to the diversity $\sigma^2(t)$, thus vanishing density or diversity imply vanishing adaptation rate $\dot{\mu}(t)$, consistently with Fisher's fundamental theorem of natural selection \cite{ewens1989interpretation}. (ii) On time scales $t > \tau_\mem$ with a convex fitness function, we have $F_2(\mu(t),\sigma^2(t);\mu_{\mem}^{(n)}(t)) > 0$ and thus $\sigma^2(t)$ decays monotonically in the absence of mutations. (iii) This emphasizes the importance of mutations $D_{\rm mut}$ to maintain a sufficient level of diversity and thus enable efficient optimization.
	Qualitatively, mutations therefore play a role similar to driving in athermal physical systems, as they break detailed balance and maintain a steady-state dynamics at large time.
	
	\begin{figure}
		\hspace*{-0.4cm}\includegraphics[]{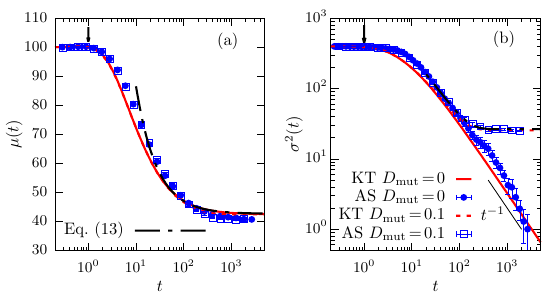}
		\caption{Adaptation of the microswimmer model to a constant target velocity $V_T=0.85$: Results of kinetic theory (KT) and agent-based simulations (AS) without mutations ($D_{\rm mut}=0$) and with mutations ($D_{\rm mut}=0.1$). The vertical arrow indicates $\tau_\mem=1$. (a) Average policy (rotation diffusion) $\mu(t)$. (b) Diversity $\sigma^2(t)$ {(variance of policy)}. The black dashed-dotted curve is Eq.~(\ref{eq:limit_Mut}) fitted to the AS data with $D_{\rm mut}=0.1$.}
		\label{fig:const_target_mod1}
	\end{figure}
	
	\emph{Phototactic, diffusive agents} -- As a first application of the above framework, we study a model of phototactic microswimmers (in $d=2$) with decentralized learning capabilities. Each swimmer moves at speed $v_0$ in a direction given by its orientation angle $\theta$ in the $(x,y)$-plane, which defines the state $\state=\theta$. As for active Brownian particles \cite{romanczuk2012active}, orientation can change randomly via rotational diffusion with strength $D_\theta,$ but also via tumbling events towards the $x-$direction (direction of the light source) \cite{martin2016}. Physical interactions between microswimmers are neglected. 
	\gj{The target behavior is a macroscopic flow along the $x$-axis in which the ensemble-averaged velocity $\langle V_x \rangle = N^{-1}\sum_{i=1}^N V_{x,i}$ is equal to the target velocity $ V_T $. This behavior can be promoted via a decentralized learning procedure in which each agent measures its own velocity $G(\state)=V_x=v_0\cos\theta$ and adapts its rotational diffusion coefficient $\policy=D_\theta$ (i.e., its policy). By maximizing the reward function, $\reward(\mem) = - (\mem-V_T)^2$, each agent's time-averaged local velocity, $\mem,$ converges towards $V_T$ and thus induces the same ensemble-averaged velocity.}
	Importantly, although physical interactions are neglected, learning is collective \gj{due to information exchange.}
	The asymptotic solutions of Eqs.~(\ref{eq:evol:mu}) and (\ref{eq:evol:sigma}) (detailed derivations in \cite{suppmat}) read, 
	\begin{align}
	\sigma^2(t) &= \sqrt{\frac{2 D_{\rm mut}}{\lambda_0}} \frac{1}{\tanh(\sqrt{2 D_{\rm mut} \lambda_0} (\tau_0 + t) )} \label{eq:limit_Mut}\\
	\mu(t) &= D_{\theta,T} {+} \frac{\eta_0 \sqrt{2 D_{\rm mut} \lambda_0}}{\sinh(\sqrt{2 D_{\rm mut} \lambda_0} (\tau_0 + t) )}, \label{eq:limit_Mut_mu}
	\end{align}
	which boil down to $\sigma^2(t)= [\lambda_0 (\tau_0 +t)]^{-1}$ and $\mu(t) = D_{\theta,T} {+} \eta_0/(\tau_0+t)$ for $D_{\rm mut} \to 0$.
	The parameter $\lambda_0= 4 \lambda_T \alpha_T \rho {V_*^\prime}^2$ is a characteristic rate,
	where
	$V_*^\prime={\partial  \bar{V}_x(D_\theta) }/{\partial D_\theta}|_{D_\theta=D_{\theta,*}}$
	is the response of the average velocity to policy changes \gj{and $\bar{V}_x(D_\theta)=\langle V_x\rangle_{D_\theta}$ is the ensemble-averaged flow for fixed policy $D_\theta.$}
	The parameters $\tau_0$ and $\eta_0$ depend on initial conditions, and $D_{\theta,*}=\policy_*$ is the predefined policy used in Eq.~(\ref{eq:mu:mem}).
	When $D_{\rm mut}>0$, $\sigma^2(t)$ converges to a plateau value $\sigma^2_{\infty}=\sqrt{2D_{\rm mut}/\lambda_0}$ when $t\to \infty$,
	while $\mu(t)$ converges to the optimal policy $D_{\theta,T}$.
	The convergence of both $\sigma^2(t)$ and $\mu(t)$ occurs over the learning time $\tau_L=(2 D_{\rm mut} \lambda_0)^{-1/2}$.
	Importantly, our theoretical analysis reveals two important control parameters: the rate $\lambda_0$ and the mutation strength $D_{\rm mut}$.
	The effect of the mutation rate $D_{\rm mut}$ is characterized by the `uncertainty relation' $\tau_L \sigma^2_{\infty} = \lambda_0^{-1}$,
	where both $\tau_L$ and $\sigma^2_{\infty}$ depend on $D_{\rm mut}$, while $\lambda_0$ is independent of $D_{\rm mut}$. The uncertainty relation states that quick convergence and adaptability (i.e., a small $\tau_L$) obtained by increasing the mutation rate $D_{\rm mut}$ comes at the cost of having large fluctuations around the target policy (i.e., a large $\sigma^2_{\infty}$), in turn implying fluctuations around the collective goal, since $\mathrm{Var}(N^{-1}\sum_{i=1}^N V_{x,i}) \propto \sigma^2_{\infty}/N$.
	Predictions of the kinetic theory are successfully compared to agent-based simulations in Fig.~\ref{fig:const_target_mod1}. Additionally,
	Eq.~(\ref{eq:limit_Mut}) can be used to fit the long-time behavior of $\sigma^2(t)$ on agent-based simulations data (see Fig.~\ref{fig:const_target_mod1}) to determine
	$\lambda_0$ and $D_{\rm mut}$, as on experimental data where such parameters are unknown.
	We extract $D_{\rm mut}^{\rm fit} = 0.094 \pm 0.002$ and $\lambda_0^{\rm fit} = (2.65 \pm 0.01) \cdot 10^{-4}$, which compare well to the agent-based model values $D_{\rm mut}=0.1$ and
	$\lambda_0=2.655\cdot 10^{-4}$. 
	
	\begin{figure}
		\hspace*{-0.2cm}\includegraphics[scale=0.8]{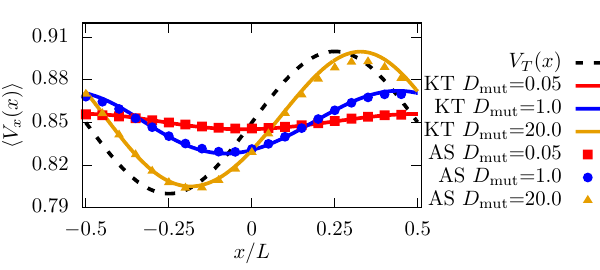}
		\caption{Adaptation of the microswimmer model to a space-dependent target $V_T(x)=0.85+ 0.05 \sin(2\pi x/L)$ (black-dashed line). Results obtained by kinetic theory (KT) and agent-based simulations  (AS) are shown for different mutation rates $D_{\rm mut}.$}
		\label{fig:space_target_mod1}
	\end{figure}

	We now consider a space-dependent target velocity $V_T(x) = V_T^s + V^\Delta_T \sin(2 \pi x / L)$ in a system of size $L$ with periodic boundary conditions.
	The reward function thus also becomes space-dependent, $\reward(\mem,x) = - \big(\mem-V_T(x)\big)^2$.
	At leading order, the steady-state average policy $\mu(x)$ satisfies \cite{suppmat},
	\begin{align}
	- V_T^s \nabla \mu(x) - \sqrt{2 D_{\rm mut} \lambda_0} (\mu(x) - D_{\theta,T}(x))=0,
	\end{align}
	where $D_{\theta,T}(x)$ is the optimal policy maximizing the local reward.
	We find for the average local velocity 
	\begin{equation}
	\langle V_x(x) \rangle=V_T^s + V^\Delta_x(\phi) \sin (2 \pi x/L - \phi),
	\end{equation}
	with a phase $\phi$ given by $\tan(\phi) = 2 \pi V_T^s / ( \sqrt{2 D_{\rm mut} \lambda_0} L)$ and an amplitude $V^\Delta_x(\phi) = V^\Delta_T \cos(\phi)$. \gj{The theory thus predicts that agents with smaller mutation rates have a larger phase $\phi$ and therefore the observed deviation from the target profile $\langle V_x(x) \rangle - V_T(x) = V_T^\Delta \big( \cos (\phi) \sin(2 \pi x/L - \phi) - \sin(2 \pi x/L)  \big) $ increases.} 
	In Fig.~\ref{fig:space_target_mod1}, we compare this predicted behavior by kinetic theory to agent-based simulations and find very good agreement for both the mutation-rate dependent amplitude $V^\Delta_x(D_{\rm mut})$ and the phase $\phi(D_{\rm mut})$. \gj{The observed phenomenology can be explained by comparing the typical advection time $\tau_A = L/V_T^s$ to the learning time $\tau_L.$ Agents with small mutation rates have $\tau_L \gg \tau_A$ thus they learn too slow and are unable to adapt in time to the environment. } Our results therefore emphasize the importance of mutations to maintain diversity and thus being able to adapt to space- or time-dependent targets.

	\begin{figure}
		\hspace*{-0.4cm}\includegraphics[]{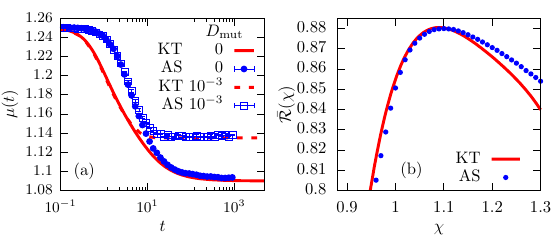}
		\caption{Adaptation of the light-sensing robot model to maximize collected light: Kinetic theory (KT) and agent-based simulations (AS) without mutations ($D_{\rm mut}=0$) and with mutations ($D_{\rm mut}=10^{-3}$). (a) Average policy $\mu(t)=\langle \chi\rangle$; (b) Average reward $\bar{\reward}(\chi)$ for robots with a given policy $\chi$.}
		\label{fig:const_target_mod2}
	\end{figure}
	
	\emph{Light-sensing robots} -- We also study a model featuring robots (in $d=1$) which attempt to maximize light collected from a heterogeneous light field $I(x)$.
	The state of a robot is $\state=(I,v)$, where $I$ is the local light intensity measured by a sensor, and $v$ is the robot velocity (diffusion is neglected, $D_0=0$).
	The memory $\mem$ averages the measured light intensity according to Eq.~(\ref{eq:mem:dyn}), with $G(\state)=I$. A state-dependent controller adapts the robot's speed
	$v(I) = v_0 - \chi I$, keeping $v(I) \geq v_{\rm min}$.  The parameter $\chi$ thus defines the sensitivity with which robots react to the external light. Robots adapt their policy $\policy=\chi$ to maximize collected light, using a reward $\reward(\mem) = \mem$.
	Fig.~\ref{fig:const_target_mod2}(a) shows that the theory successfully describes the optimization of $\chi$,
	by predicting the long-time behavior of $\mu(t)$ (calculations in SM \cite{suppmat}).
	In the absence of mutations ($D_{\rm mut}=0$), the diversity $\sigma^2(t)$ goes to zero and $\mu(t)$ approaches the maximal reward around $\chi_T=1.09$ [see Fig.~\ref{fig:const_target_mod2}(b)]. However, for $D_{\rm mut}>0$, mutations cause the robots to learn a distribution of policies with $\sigma^2_{\infty} > 0$. Importantly, the underlying model implies a pronounced asymmetry $\bar{\reward}(\chi_T + \epsilon) > \bar{\reward}(\chi_T - \epsilon), \epsilon>0$ in the reward function [Fig.~\ref{fig:const_target_mod2}(b)].
	\gj{Here, $\bar{\reward}(\chi)\coloneqq\reward(\mu_\mem(\chi))$ is the average steady-state reward for an agent with policy $\chi$, which we can derive exactly since the assumed time scale separation implies the existence of a direct relationship between the policy and the steady-state average memory, $\mu_\mem(\chi)$}.
	This asymmetry makes it preferable to adapt for $t\to \infty$ to $\mu_{\infty} > \chi_T$ in the presence of mutations, as observed in the agent-based simulations.
	Expanding $\mu_\mem$ to fourth order in Eq.~(\ref{eq:mu:mem}),
	KT provides a quantitative prediction of $\mu_{\infty}$ [Fig.~\ref{fig:const_target_mod2}(a)]. \gj{The increased steepness of $\bar{\reward}(\chi)$ for $\chi > 1.2$ in KT implies a larger $V_*^\prime = \partial \bar{\reward}(\chi) / \partial \chi$ and thus explains the faster adaptation speed observed for KT in Fig.~\ref{fig:const_target_mod2}(a).}

	\emph{Conclusion and outlook} -- We proposed a kinetic theory framework for decentralized learning in smart active matter, in which agents maximize their reward $\reward$ by learning improved policies $\policy$ from neighboring agents, to reach a target collective \gj{behavior}. Our theory enables derivation of space-dependent field equations describing the time-dependence of policy distributions. \gj{ Assumptions required to derive closed equations include weak correlations between agents, a hierarchy of time scales, a Gaussian approximation for the $\mem$- and $\policy$-dependence, and expansions of the teaching probability $p_T$ and the average memory. Most of these assumptions can be lifted for specific applications, however, the theory cannot quantitatively account for many-body interactions and correlations.    }
	We have applied the framework to two explicit models, and predictions favorably compared to numerical simulations of the agent-based model, showing a rich phenomenology. 
	Future research may include the analysis of learning with different interaction ranges, multi-dimensional \gj{or discrete} policies and more complex reward functions \gj{such as time-dependent or non-monotonic functions or multiple species with competing reward. More generally, it will be crucial to explore systematic techniques guiding the design of reward functions facilitating targeted collective behavior. Addressing this inverse problem may be possible by incorporating higher-level learning mechanisms which are able to iterativelly optimize the agent-based reward function itself.} 
	While we considered for simplicity a limit of separated time scales between physical, memory and learning dynamics, our framework allows for systematic improvements over this approximation, notably taking into account the effect of the communication range. In addition, it would also be of interest to investigate the generality of the uncertainty relation linking learning time and policy fluctuations.

	
	\acknowledgments  
	
	\emph{Acknowledgments} -- The authors are grateful to Olivier Dauchot and Bahram Houchmandzadeh for stimulating discussions.
	This work has been supported by a grant from MIAI@Grenoble Alpes and the Agence Nationale de la Recherche under France 2030 with the reference ANR-23-IACL-0006.

	\bibliography{library_local.bib}
	
	
	
	\vspace{1.5cm}
	
	\paragraph{End Matter} -- We describe here the general framework of kinetic theory of decentralized learning introduced in this work. Calculation details for specific models are reported in the SM~\cite{suppmat}.
	
	The term $\gj{\mathcal{I}_{\rm teach}[f]}$, introduced in Eq.~(\ref{eq:KT:policy}), can be formulated in a rather similar way as usual collision terms in kinetic theories.
	It is convenient to express $\mathcal{I}_{\rm teach}[f]$ in a similar way as a master equation for a stochastic jump process, using a $f$-dependent effective transition rate
	$W_f(\mem',\policy'|\mem,\policy; \bm{x},t)$ defined as, 
	\begin{align}
	\hspace{-0.5cm} \nonumber 
	&W_f(\mem',\policy'|\mem,\policy; \bm{x},t)\\ \nonumber
	&\qquad\quad = 2 \lambda_T \hspace{-0.1cm} \int \hspace{-0.1cm} d\state_2 \int \hspace{-0.1cm} d\mem_2 \hspace{-0.1cm}\int \hspace{-0.1cm} d \policy_2 \int \hspace{-0.1cm} d\bm{x}_2 K(\bm{x}_2,\bm{x})\\ \nonumber
	& \qquad\qquad \times p_T(\reward(\mem_2),\reward(\mem)) f(\state_2,\mem_2,\policy_2,\bm{x}_2,t) \nonumber \\
	&\qquad\qquad \times \delta(\policy^\prime-\policy_{2}) \delta(\mem^\prime-\mem_2).
	\label{eq:def:WL}
	\end{align}
	The transition rate $W_f$ encodes the microscopic learning dynamics of the agent model: the probability that a focus particle 1, at position $\bm{x}$ and characterized by $(\state,\mem,\policy)$, 
	learns its memory $\mem'=\mem_2$ and policy $\policy'=\policy_2$ from a particle 2 at position $\bm{x}_2$ and with $(\state_2,\mem_2,\policy_2)$, is proportional to the teaching rate $\lambda_T$, the distance-dependent kernel
	$K(\bm{x}_2,\bm{x})$, and to the teaching probability $p_T(\reward(\mem_2),\reward(\mem))$ (i.e., the probability that particle 2 becomes the teacher of particle 1),
	defined in Eq.~(\ref{eq:def:PT}).
	The kernel $K(\bm{x}_2,\bm{x})$ has a characteristic range $\delta_c$, e.g., $K(\bm{x}_2,\bm{x})=(2\pi\delta_c^2)^{-1} \exp[-(\bm{x}_2-\bm{x})^2/2\delta_c^2]$ in 2D, and is normalized to $1$.
	Note that if the agent learns the policy $\policy_2$ it also adapts its memory to the value $\mem_2$ in order to evaluate the correct reward associated with policy $\policy_2$. \gj{This choice is motivated by Ref.~\cite{ben2023morphological} which have similarly exchanged the information about the reward function sampled by the robots to avoid additional long-time sampling. By replacing $\delta(\mem^\prime-\mem_2) \rightarrow \delta(\mem^\prime-\mem)$, the student would maintain their previous memory, which might be a more intuitive choice in other applications. }
	The state $\state$ of particle 1 is not included in $W_f(\mem',\policy'|\mem,\policy; \bm{x},t)$ because $\state$ is not modified by the learning process.
	The factor of $2$ in front of $\lambda_T$ accounts for the fact that during one adaptation step either agent 1 is the teacher and agent 2 the student or the other way around.
	
	Note also that at variance with usual kinetic theories (e.g., of gases, granular gases or active matter), the occurrence of learning interactions is not controlled by the kinetics of particle collisions,
	but rather by an imposed \gj{teaching} rate $\lambda_T$, which in practice may be much smaller than the rate of particle encounter.
	We thus assume that information propagation due to the sole communication range, at a speed at most at the order of $\lambda_T \delta_c$, remains much slower than information \gj{transport} by particle motion
	at a typical speed $v_0$ (i.e., we assume $\lambda_T \delta_c \ll v_0$).
	This means that we can neglect the distinction between $\bm{x}_2$ and $\bm{x}$ in Eq.~(\ref{eq:def:WL}), thereby yielding a purely local expression of $W_f$,
	\begin{equation}
	\label{eq:def:WL2}
	W_f(\mem',\policy'|\mem,\policy)
	= 2 \lambda_T p_T(\reward(\mem'),\reward(\mem)) \fzero(\mem',\policy'),
	\end{equation}
	where we have introduced the marginal distribution,
	\begin{equation}
	\fzero(\mem,\policy) = \int d \state f(\state,\mem,\policy).
	\end{equation}
	Here, and in the following, we drop the explicit dependence on $\bm{x}$ and $t$ for the sake of brevity. 
	\gj{Within kinetic theory} the \gj{teaching} term $\mathcal{I}_{\rm teach}[f]$ is formally obtained from an effective master equation, and takes the form,
	\begin{align}
	\mathcal{I}_{\rm teach}[f] &= \int \hspace{-0.1cm} d \mem^\prime \hspace{-0.1cm} \int  \hspace{-0.1cm} d \policy^\prime \left[ W_f(\mem,\policy |\mem',\policy') f(\state,\mem^\prime,\policy^\prime) \right. \nonumber \\ 
	& \qquad \qquad \left. - W_f(\mem',\policy'|\mem,\policy) f(\state,\mem,\policy) \right].
	\label{eq:MEq:kin:theo}
	\end{align}
	By inserting the explicit form of the transition rate in Eq.~(\ref{eq:def:WL2}), the \gj{teaching} term can be written down explicitly,
	\begin{align}
	&\gj{\mathcal{I}_{\rm teach}[f]} = \\
	& \quad 2\lambda_T f_0(\mem,\policy)   \int d \mem^\prime p_T(\reward,\reward^\prime)\int d \policy^\prime f(\state,\mem^\prime,\policy^\prime) \nonumber	\\
	& \quad - 2\lambda_T f(\state,\mem,\policy)  \int d \mem^\prime    p_T(\reward^\prime,\reward) \int d \policy^\prime \fzero(\mem^\prime,\policy^\prime),  \nonumber
	\end{align}
	with the shorthand notations $\reward=\reward(\mem)$ and $\reward'=\reward(\mem')$.
	Writing decentralized learning in such a general way and thus introducing a framework for the systematic derivation of macroscopic hydrodynamic equations for population dynamics is one of the major results of this manuscript. \gj{As in usual kinetic theories, we have assumed that the two-body phase-space density can be approximated by multiplying the one-body phase-space densities, thus assuming weak correlations between individual agents. } In the following, we now aim at transforming these equations to derive macroscopic equations for the dynamics of policies.
	
	In a first reduction step, we integrate out the dependence of $f$ on the state $\state$ and memory $\mem$ to focus on the macroscopic dynamics of policies. As a first approximation this can be achieved by assuming $\tau_G \ll \tau_\mem \ll \tau_T$, i.e., that the time scale $\tau_\mem$ is large enough \gj{to smooth out} 
	the fluctuations in the observations of $G(\state)$ on a time scale $\tau_G$, 
	and that the learning occurs on time scales much larger than changes in the memory. \gj{This assumption also implies that exchanging information on the memory during learning does not impact the reported results. We believe this assumption corresponds to a natural learning scenario, allowing each agent to sample as much information as possible before learning. However, the assumption of a hierarchy in time scales is not a fundamental limitation of the theory and can also be lifted if this is desirable for specific applications.}
	
	When studying the time-dependence of $f$ on a time scale $\tau_\mem$, the physical interaction terms in $\mathcal{I}_{\rm phys}[f]$ \gj{drop out (leaving only the diffusive and convective terms)}. \gj{Transforming the agent-based memory dynamics in Eq.~(\ref{eq:mem:dyn}) into an evolution equation for the probability distribution of $\mem$ \cite{risken1996fokker} leads to} the explicit relation for the memory dynamics,
	\begin{equation}
	I_{\rm mem}[f] =  \tau_\mem^{-1} \Big(f(\state,\mem,\policy) + (\mem - G(\state)) \frac{\partial f(\state,\mem,\policy)}{ \partial \mem} \Big),
	\end{equation}
	we find a closed equation for the time-dependence of the marginal distribution,
	\begin{align}
	\frac{\partial f_0(\mem,\policy)}{\partial t} &= - \nabla J_0 + \tau_\mem^{-1} \Big(f_0 + (\mem - \bar{G}(\policy)) \frac{\partial f_0}{ \partial \mem} \Big) \nonumber\\
	+& 2\lambda_T f_0  \hspace{-0.1cm} \int \hspace{-0.1cm} d \mem^\prime \tanh(\alpha_T ( \reward -\reward^\prime )) \hspace{-0.1cm}\int \hspace{-0.1cm} d \policy^\prime \fzero(\mem^\prime,\policy^\prime) \nonumber\\
	+&D_{\rm mut} \frac{\partial^2}{\partial \policy^2} f_0 \label{eq:marginal_end_matter}
	\end{align}
	with the current $J_0 = \Big (  \bar{V}(\policy)  - D_0 \nabla \Big) f_0  $. At this level, \gj{due to the time scale separation}, the details of the microscopic model thus only enter via the policy-dependent average of the observed information $ \bar{G}(\policy)\coloneqq \langle G(\state) \rangle_\policy,$ and the policy-dependent average velocity $ \bar{V}(\policy)\coloneqq \langle V \rangle_\policy.$ 	The brackets $\langle...\rangle_\policy$ denote the policy-dependent ensemble average used as a proxy for the running average on a time scale $\tau_\mem$ due to the assumed time scale separation.

	As in the main text, we then introduce the marginal phase-space distribution of policies [Eq.~(\ref{eq:phi0})], the agent density $\rho(\bm{x},t)=\int d\policy \varphi_0(\policy,\bm{x},t)$ and the average memory, $\mu_\mem(\policy,\bm{x},t) = \varphi_0^{-1} \int d \mem \mem f_0(\mem,\policy,\bm{x},t) $ [Eq.~(\ref{eq:mem})]. \gj{Due to the assumed time scale separation a Gaussian dependence of $f_0$ on $\mem$ can be justified with the central limit theorem.}  To find closed equations we approximate the activation function to first order, $\tanh(\alpha_T ( \reward -\reward^\prime )) = \alpha_T ( \reward -\reward^\prime ) + \mathcal{O}(\alpha_T^3 ) $. \gj{This limits applications of the theory to cases where it is not necessary to evaluate the reward with very high precision, and learning is achieved as long as sufficient teaching events are performed.}  After some calculations we find,
	\begin{align}
	\frac{\partial \varphi_0(\policy)}{\partial t} &= - \nabla J + D_{\rm mut} \frac{\partial^2}{\partial \policy^2} \varphi_0(\policy) \\&+ \lamLt  \rho  \int d \mem f_0(\mem,\policy) \reward(\mem)   \nonumber\\
	&   - \lamLt  \varphi_0(\policy) \int d \policy^\prime \int d \mem^\prime \fzero(\mem^\prime,\policy^\prime) \reward(\mem^\prime), \nonumber
	\end{align}
	with the policy current $J = \Big (  \bar{V}(\policy)  - D_0 \nabla \Big) \varphi_0  $.  We also introduce the renormalized teaching rate, $\lamLt= 2 \lambda_T \alpha_T$.  For the average memory we can derive,
	\begin{align}
	\frac{\partial \mu_\mem(\policy)}{\partial t}&=  \tau_\mem^{-1} \Big(\bar{G}(\policy) - \mu_\mem(\policy)\Big).
	\end{align}
	\gj{This equation would also include terms proportional to $D_0,$ $D_{\rm mut}$ and $\lamLt$, which, however, are negligible compared to $\tau_\mem^{-1}$ within the approximations applied in the present manuscript.}
	
	\gj{Subsequently, we perform the reduction procedure as described in the main text to derive closed equations for the average policy $\mu(t) $ and the diversity $\sigma^2(t),$ as presented in Eq.~(\ref{eq:evol:mu}). The closure is based on the assumption of a Gaussian dependence of the reduced phase-space density on $\policy.$ This assumption is reasonable for the examples discussed in the present manuscript featuring an unbounded policy space and convex fitness functions. (The phototactic agents had policies $\mu(t) \gg \sigma(t)$, thus this bound did not intervene with the results.) If Gaussianity cannot be assumed, for example in applications in which the policy space is bounded, other distributions such as the $\beta-$distribution or multimodal distributions may be considered. }\\
	
	\gj{ While we have focused on continuous quantities within this manuscript, our theory can certainly be adapted to accommodate discrete quantities. This includes, in particular, discrete policies which might be relevant for decision-making. 
		Let $\policy_0$ be the set of possible discrete policies, and $f_{\policy,0}(\mem,\bm{x},t)$ be the phase-space density of agents with policy $\policy \in \policy_0$. Then Eq.~(\ref{eq:marginal_end_matter}) becomes,
		\begin{align}
		\frac{\partial f_{\policy,0}(\mem)}{\partial t} &= - \nabla J_0 + \tau_\mem^{-1} \Big(f_0 + (\mem - \bar{G}_\policy) \frac{\partial f_{\policy,0}}{ \partial \mem} \Big) \nonumber\\
		+& 2\lambda_T f_0  \hspace{-0.1cm} \int \hspace{-0.1cm} d \mem^\prime \tanh(\alpha_T ( \reward -\reward^\prime )) \sum_{\policy^\prime \in \policy_0} f_{\policy^\prime,0}(\mem^\prime) \nonumber\\
		+&\sum_{\policy^\prime \in \policy_0} \lambda_{\rm mut, \policy^\prime \rightarrow \policy}  f_{\policy^\prime,0} \label{eq:marginal_end_matter_discrete}.
		\end{align}
		The only differences to Eq.~(\ref{eq:marginal_end_matter})  are the replacement of the integral by a sum, and introducing a discrete master equation describing mutation via a random jump process in policy space.
	}

\end{document}


\setlength{\belowdisplayskip}{3pt} \setlength{\belowdisplayshortskip}{3pt}
	\setlength{\abovedisplayskip}{3pt} \setlength{\abovedisplayshortskip}{3pt}
	
	\title{Supplemental Material for ``Kinetic theory of decentralized learning for smart active matter''}
	
	\author{Gerhard Jung}

	\author{Misaki Ozawa}

	\author{Eric Bertin}

	\date{\today}
	
	\maketitle
	
	\setcounter{equation}{0}
	\setcounter{figure}{0}
	\setcounter{table}{0}
	\setcounter{page}{1}
	\renewcommand{\theequation}{S\arabic{equation}}
	\renewcommand{\thefigure}{S\arabic{figure}}
	\renewcommand{\bibnumfmt}[1]{[S#1]}
	\renewcommand{\citenumfont}[1]{S#1}
	
	In this supplemental material (SM) we provide details on the theoretical derivation of the kinetic theory. This includes theoretical details for the microswimmer model (Sec.~\ref{sec:mod1}) and the light-sensing robot model (Sec.~\ref{sec:mod2}), and the derivation of the multi-dimensional case (Sec.~\ref{sec:multidim}).
	
	\section{Derivations for MODEL1: Microswimmers}
	\label{sec:mod1}
	
	\begin{figure}
		\includegraphics{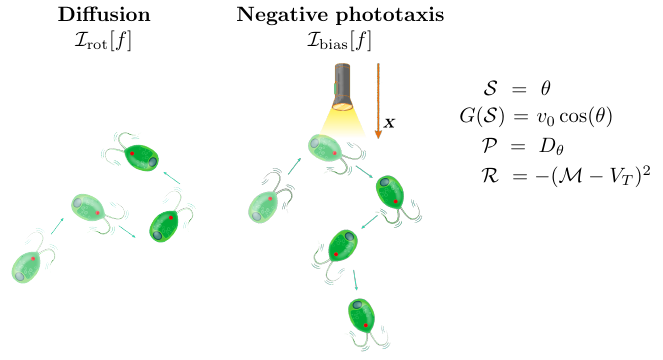}
		\caption{Illustration of the two physical interaction terms. $\mathcal{I}_{\rm rot}[f]$ just leads to random reorientation, while $\mathcal{I}_{\rm bias}[f]$ describes (negative) phototaxis and thus biases the motion of the microswimmer towards the positive $x-$direction. }
		\label{fig:mod1}
	\end{figure}
	
	The first model we analyze consists of non-interacting microswimmers in two-dimensions. Each swimmer is characterized by its individual rotational diffusion coefficient $D_\theta=\policy$ which will also correspond to the policy we are aiming to optimize. Furthermore, the microswimmers tend to align into the direction of an external light source in the $x-$direction (see Fig.~\ref{fig:mod1}).  In the model, the speed $v_0$ of the swimmer is constant, $\bm{v} = v_0 \bm{n} $ and thus only the heading vector $\bm{n}= (\cos \theta, \sin \theta)$ is fluctuating. Here, $\theta=\state$ is therefore the internal variable of interest. The interaction term then reads,
	\begin{align}
	\mathcal{I}_{\rm phys}[f] &= \gj{-\nabla \cdot \left( \bm{v}  f \right) + D_0 \Delta f} + \mathcal{I}_{\rm rot}[f] + \mathcal{I}_{\rm bias}[f],\\
	\mathcal{I}_{\rm rot}[f] &= D_\theta \frac{\partial^2}{\partial \theta^2}f ,   \\
	\mathcal{I}_{\rm bias}[f] &= -\lambda_B f + \lambda_B P_B(\theta ) f_0.& \label{eq:tumble}
	\end{align}
	Here, \gj{$\nabla = (\partial/\partial x,\partial/\partial y),$ $\Delta = \nabla \cdot \nabla$ is the Laplace operator, $D_0$ is the space diffusion coefficient,  $f_0$ is the marginal distribution defined in Eq.~(18)} and $\lambda_B$ is the rate with which the microswimmers tumble and thus bias their orientation $\theta$ towards the distribution $P_B(\theta)$, \gj{centered around} the $x$-axis. \gj{This allows us to model the phototactic behavior of the microswimmers.}   We assume that $P_B$ is Gaussian with the standard deviation $\sigma_B.$ \gj{The mathematical form in Eq.~(\ref{eq:tumble}) is comprised of a source term $+ \lambda_B P_B(\theta ) f_0$ describing the increase in probability of finding an agent with orientation $\theta$ after the tumbling, and a sink term $-\lambda_B f$ describing those agents which had an orientation $\theta$ before the tumbling. In agent-based simulations, these tumbling events are modeled by iterating over all agents and letting them perform a tumbling event with the rate $\lambda_B.$ In a tumbling rate, a new orientation $\theta_{\rm new}$ is assigned to the agents. The orientation $\theta_{\rm new}$ is independent of the current orientation and drawn randomly from $P_B(\theta).$  }
	
	In consequence, for a given bias strength, the microswimmer can adapt its average velocity $\langle V_x \rangle$ in $x-$direction \gj{(mathematical definition see below)}  by choosing a larger \gj{rotational} diffusion coefficient (small $\langle V_x \rangle$) or a smaller \gj{rotational} diffusion coefficient (large $\langle V_x \rangle$). This model is inspired by the behaviour of the microalga Chlamydomonas Reinhardtii which features negative phototaxis, a very similar model has already been analyzed theoretically \cite{bertin2019micro,bertin2020micro}.
	
	\subsection{Solving the microscopic physical model}
	
	\gj{In the absence of teaching and mutations, considering only the $\mathcal{I}_{\rm phys}$ dynamics, we can collect the above terms and the hydrodynamic equations for the above model can thus be written as, 
		\begin{equation}
		\frac{\partial f(\theta,\bm{x},t)}{\partial t}  = -\nabla \cdot \left( \bm{v}  f(\theta,\bm{x},t) \right) + D_0 \Delta f(\theta,\bm{x},t) + D_\theta \frac{\partial^2}{\partial \theta^2}f(\theta,\bm{x},t) -\lambda_B f(\theta,\bm{x},t) + \lambda_B P_B(\theta ) f_0(\bm{x},t).
		\end{equation}
		This equation includes in order from left to right: convection, diffusion in space, diffusion in orientation and tumbling/phototaxis.} Since the model is two-dimensional, to simplify calculations we introduce the complex vector notation, $n = n_x + \mathrm{i} n_y = e^{\mathrm{i} \theta}.$ Therefore $n_x = \text{Re}(n)$ and $n_y = \text{Im}(n)$. To remove the dependence of the model on the orientation $\theta$ we expand $f$ using its Fourier-transform, $f_k = \int_{-\pi}^{\pi} d\theta e^{\textrm{i} \theta k} f(\theta)$. Applying the same methodology as discussed in Refs.~\cite{bertin2019micro,peshkov2014boltzmann} we can thus derive the time-dependence of the distribution $f_k$ without the contribution of memory and teaching,
	\begin{equation}
	\frac{\partial f_k}{\partial t} + \frac{v_0}{2} (\nabla f_{k-1} + \nabla^* f_{k+1}) - D_0 \Delta f_k = - D_\theta k^2 f_k - \lambda_B (f_k - f_0 e^{-\sigma_{B}^2 k^2 / 2} ). 
	\end{equation}
	\gj{Here, $\nabla = \frac{\partial}{\partial x} + {\rm i}\frac{\partial}{\partial y}$ and $\nabla^* = \frac{\partial}{\partial x} - {\rm i}\frac{\partial}{\partial y}$.}
	
	\gj{To derive the full hydrodynamic equations with teaching, we now additionally include the novel terms introduced in the main manuscript, i.e., $\mathcal{I}_{\rm mem}$, $\mathcal{I}_{\rm teach}$, and $\mathcal{I}_{\rm mut}$.} For the memory term, $\mathcal{I}_{\rm mem}[f]$, we choose $G(\mathcal{S})=v_0 \cos \theta$, corresponding to the instantaneous velocity in $x-$direction. \gj{Starting from Eq.~(21) in the main manuscript, we thus find for the modes of the memory dynamics $\mathcal{I}_{\rm mem}$,}
	\begin{align}
	\int_{-\pi}^{\pi} d\theta  \mathcal{I}_{\rm mem}[f] &= \tau_\mem^{-1}   \int_{-\pi}^{\pi} d\theta  \Big(f + (\mem - v_0 \cos(\theta)) \frac{\partial f(\theta,\mem,\policy,\bm{x},t)}{ \partial \mem} \Big)\\
	&= \tau_\mem^{-1}  \Big(f_0 + \Big(\mem \frac{\partial f_0}{ \partial \mem} - v_0 \frac{\partial \text{Re}(f_1)}{\partial \mem}  \Big) \gj{\Big)}\\
	\int_{-\pi}^{\pi} d\theta e^{\textrm{i} \theta}  \mathcal{I}_{\rm mem}[f] &= \tau_\mem^{-1}  \Big(f_1 + \Big(\mem \frac{\partial f_1}{ \partial \mem} - v_0 \frac{\partial \hat{f_2}}{\partial \mem}  \Big) \gj{\Big)}
	\end{align}
	where we have defined $\hat{f_2} = \int_{-\pi}^{\pi} d\theta e^{\textrm{i} \theta} \cos(\theta) f(\theta,\mem,\policy,\bm{x},t) = \frac{1}{2} (f_0 + f_2).$
	\gj{Similarly, we can calculate the modes of $\mathcal{I}_{\rm teach}$ (starting from Eq.~(20)) and $\mathcal{I}_{\rm mut},$ 
		\begin{align}
		\int_{-\pi}^{\pi} d\theta  \mathcal{I}_{\rm teach}[f] &=  2\lambda_T f_0    \int d \mem^\prime \tanh(\alpha_T ( \reward -\reward^\prime ))\int d \policy^\prime \fzero(\mem^\prime,\policy^\prime),\\
		\int_{-\pi}^{\pi} d\theta e^{\textrm{i} \theta}  \mathcal{I}_{\rm teach}[f] &=  2\lambda_T f_0(\mem,\policy)   \int d \mem^\prime p_T(\reward,\reward^\prime)\int d \policy^\prime f_1(\mem^\prime,\policy^\prime)  - 2\lambda_T f_1(\mem,\policy)  \int d \mem^\prime    p_T(\reward^\prime,\reward) \int d \policy^\prime \fzero(\mem^\prime,\policy^\prime), \\
		\int_{-\pi}^{\pi} d\theta e^{k \textrm{i} \theta}  \mathcal{I}_{\rm mut}[f] &= D_{\rm mut}  \frac{\partial^2}{\partial \policy^2} f_k,
		\end{align}
		where \gj{$\lambda_T$ is the rate of the teaching events and $D_{\rm mut}$ the mutation rate}. Subsequently, we can insert $
		p_T(\reward,\reward^\prime) = \frac{1}{2} \left( 1+ \tanh\big[\alpha_T \big(\reward - \reward^\prime \big)\big] \right).$
	}
	
	We are interested in the dynamics of the two lowest \gj{modes}, the marginal distribution $f_0$ corresponding to the density and the first-order \gj{mode} $f_1,$ which is related to the average velocity, 
	\gj{
		\begin{equation}
		\langle {\bm V} \rangle_\policy = \frac{\int_{-\pi}^{\pi} d\theta f(\theta,\mem,\policy,\bm{x},t) v_0 {\bm n}}{\int_{-\pi}^{\pi} d\theta f(\theta,\mem,\policy,\bm{x},t)} .
		\end{equation}
		Since $f(\theta,\mem,\policy,\bm{x},t)$ is an even function in terms of $\theta$ in our setting, $\langle {\bm V} \rangle_\policy=(\langle V_x \rangle_\policy, 0)$ (the $y-$component is zero), where $\langle V_x \rangle_\policy$ is the $x-$component, given by
		\begin{equation}
		\langle V_x \rangle_\policy = v_0 \frac{\text{Re}(f_1(\mem,\policy,\bm{x},t))}{f_0(\mem,\policy,\bm{x},t)} .
		\end{equation}
	}
	
	For the two lowest \gj{modes we therefore finally find for the full hydrodynamic equations including teaching by collecting all terms derived above and },
	\begin{align}
	\frac{\partial f_0(\mem,\policy,\bm{x},t)}{\partial t}  &= - v_0 \text{Re}(\nabla^* f_1) + D_0 \Delta f_0 + \tau_\mem^{-1} \Big(f_0 + \Big(\mem \frac{\partial f_0}{ \partial \mem} - v_0\gj{\frac{\partial \text{Re}(f_1)}{\partial \mem}}  \Big)\Big) \nonumber\\
	&+ D_{\rm mut} \frac{\partial^2}{\partial \policy^2} f_0 + 2\lambda_T f_0    \int d \mem^\prime \tanh(\alpha_T ( \reward -\reward^\prime ))\int d \policy^\prime \fzero^\prime  ,\\
	\frac{\partial f_1(\mem,\policy,\bm{x},t)}{\partial t}  &= - \frac{v_0}{2}  (\nabla f_0 + \nabla^* f_2) + D_0 \Delta f_1 - D_\theta f_1- \lambda_B(f_1-f_0 e^{-\sigma_B^2 / 2}) \nonumber\\
	&+\tau_\mem^{-1} \Big(f_1 + \Big(\mem \frac{\partial f_1}{ \partial \mem} - \frac{v_0}{2} \Big ( \frac{\partial {f_0}}{\partial \mem} + \frac{\partial {f_2}}{\partial \mem} \Big )  \Big)\Big) + D_{\rm mut} \frac{\partial^2}{\partial \policy^2} f_1 +  \lambda_T f_0   \int d \mem^\prime \tanh(\alpha_T ( \reward -\reward^\prime ))\int d \policy^\prime f_1^\prime \nonumber\\
	& -  \lambda_T f_1   \int d \mem^\prime \tanh(\alpha_T ( \gj{ \reward^\prime -\reward } ))\int d \policy^\prime \fzero^\prime \gj{+ \lambda_T f_0 \int d \mem^\prime \int d \policy^\prime f_1^\prime  - \lambda_T f_1 \int d \mem^\prime \int d \policy^\prime f_0^\prime}  ,
	\end{align}
	where $f_k^\prime = f_k(\mem^\prime,\policy^\prime,\bm{x},t)$.
	In the following, we will assume that the physical time scales $\lambda^{-1}_B$ and $D^{-1}_\theta$ are much smaller than $\tau_\mem$, $\lambda_T^{-1}$, $D_{\rm mut}^{-1} $ , $ \Delta L^2 / D_0 $ and $ \Delta L /v_0$, where $\Delta L$ is a typical length scale in the system. \gj{In other words, on the time scale on which $f_1$ relaxes due to rotational diffusion $D_\theta$ and bias $\lambda_B$, we can assume all other terms to not contribute to the dynamics.   } Further, we will use $\partial f_1 / \partial t$ to calculate $\partial \langle V_x(t) \rangle_\policy / \partial t$ \gj{based on the above relation $ \langle V_x \rangle_\policy = v_0 f_0^{-1} \text{Re}(f_1).$} This finally yields,
	\begin{align}
	\frac{\partial \langle V_x(t) \rangle_{D_\theta}}{\partial t}  &=  -D_\theta \langle V_x(t) \rangle_{D_\theta} - \lambda_B(\langle V_x(t) \rangle_{D_\theta}-v_0 e^{-\sigma_B^2 / 2}).
	\end{align}
	We can then solve the above equation analytically to find the steady state solution,
	\begin{equation}\label{eq:V}
	\bar{V}_x(D_\theta)=\langle V_x \rangle_{D_\theta} =  v_0\frac{ \lambda_B e^{-\sigma_B^2 / 2}}{D_\theta + \lambda_B  }.
	\end{equation}
	\gj{Here, $\langle V_x \rangle_{D_\theta}$ depends only on the policy $D_\theta$ due to the time scale separation. Importantly,} by inserting this relation into the term,  $v_0{\partial \text{Re}(f_1})/{\partial \mem}= \langle V_x \rangle_{D_\theta} \partial f_0 / \partial \mem$ we find that $\bar{V}_x(D_\theta)$ directly corresponds to the term $\bar{G}(\policy)$ as introduced in Eq.~(22) in the End Matter section of the main manuscript, i.e. the average of the information $G(\state)=V_x$ which is expected to be observed for an agent with policy $\policy.$  Importantly, one might also be able to find more complex analytical solutions, for example when there is no clear timescale separation between the physical timescales  $\lambda^{-1}_B=\tau_G$ and the memory time scale $\tau_\mem$.  Additionally, when using a more complex physical model and $\mathcal{I}_{\rm phys}[f]$ potentially features interactions between agents, one needs to make suitable approximations to find an analytical solution for $\bar{V}_x(D_\theta).$ This can be achieved by using standard approximations applied to kinetic theories \cite{peshkov2014boltzmann}.
	
	\subsection{Hydrodynamic equations}
	
	As introduced in the main manuscript, we will choose as reward function, $\reward(\mem) = - (\mem-V_T)^2.$ \gj{It should be noted that $V_T$ can be chosen as both time- and space-dependent to investigate the effect of environment changing in time or space, $V_T= V_T(\bm{x},t)$. This would not change anything on the following derivation. } The reward function enables us to expand $f_0$ in terms of its density,  $\varphi_0(D_\theta,\bm{x},t) = \int d \mem  f_0(\mem,D_\theta,\bm{x},t) $,  the average memory, $\mu_\mem(D_\theta,\bm{x},t) = \varphi_0^{-1} \int d \mem \mem f_0(\mem,D_\theta,\bm{x},t) $ and the memory variance, $\sigma^2_\mem(D_\theta,\bm{x},t) = \varphi_0^{-1} \int d \mem (\mem - \mu_\mem(D_\theta,\bm{x},t) )^2 f_0(\mem,D_\theta,\bm{x},t) $ \gj{by assuming a Gaussian distribution in $\mem$}. Finally, we can thus rewrite Eq.~(22) in the main manuscript and find,
	\begin{align}
	\frac{\partial \varphi_0(D_\theta,\bm{x},t)}{\partial t} &= - \nabla \cdot \Big ( \langle {\bm V} \rangle_{D_\theta} - D_0 \nabla \Big) \varphi_0 + \lamLt \varphi_0 \int dD_\theta^\prime\varphi_0(D_\theta^\prime,\bm{x},t) \Big( \sigma^2_\mem(D_\theta^\prime,\bm{x},t) + \mu_\mem^2(D_\theta^\prime,\bm{x},t) -2 V_T \mu_\mem(D_\theta^\prime,\bm{x},t) \Big) \nonumber
	\\&+\lamLt \varphi_0 \rho(\bm{x},t) \Big( 2 V_T  \mu_\mem - \mu_\mem^2 -\sigma^2_\mem \Big) + D_{\rm mut} \frac{\partial^2}{\partial D_\theta^2} \varphi_0 , \\
	\frac{\partial \mu_\mem(D_\theta,\bm{x},t)}{\partial t}&= - \gj{\langle {\bm V} \rangle_{D_\theta} \cdot} \nabla \mu_\mem + D_0 \Delta \mu_\mem + \frac{2D_0}{\varphi_0} \nabla \mu_\mem \gj{\cdot} \nabla \varphi_0 + \tau_\mem^{-1} \Big(\bar{V}_x(D_\theta) - \mu_\mem\Big) + 2 \lamLt \gj{\rho} \sigma_\mem^2 (V_T - \mu_\mem) \nonumber\\
	& + D_{\rm mut} \frac{\partial^2}{\partial D_\theta^2} \mu_\mem + \frac{2D_{\rm mut}}{\varphi_0} \left(\frac{\partial}{\partial D_\theta} \mu_\mem \right) \left(\frac{\partial}{\partial D_\theta} \varphi_0 \right) ,
	\end{align}
	\gj{where $\langle {\bm V} \rangle_{D_\theta}=(\langle V_x \rangle_{D_\theta}, 0)$.}
	Here, we have introduced the renormalized teaching rate, $\lamLt= 2 \lambda_T \alpha_T$.
	Applying the same timescale separation as introduced above, the space-dependence of $\mu_\mem(D_\theta,\bm{x},t)$ can be neglected and we can therefore derive the relation, 
	\begin{align}\label{eq:mem2}
	\frac{\partial \mu_\mem(D_\theta,t)}{\partial t}&=  \tau_\mem^{-1} \Big(\bar{V}_x(D_\theta) - \mu_\mem(D_\theta,t)\Big),
	\end{align}
	which corresponds to Eq.~(23) of the main manuscript. Within this approximation $\sigma^2_\mem$ becomes independent of the position and the policy, $\sigma^2_\mem(D_\theta,\bm{x},t) = \sigma^2_\mem(t)$, and thus drops out in the equation for $\varphi_0.$ Our goal is to expand $\varphi_0$ in moments of $D_\theta$ and thus derive relations for the policy dynamics described by its mean and variance. We therefore expand $\mu_\mem(\bm{x},D_\theta,t)$ around a predefined policy $D_{\theta*}=\policy_*,$ \gj{which we assume to be space independent,}
	\begin{align}
	\frac{\partial \mu^{(n)}_\mem(t)}{\partial t}&= \tau_\mem^{-1} \Big(\bar{V}_x^{(n)} - \mu^{(n)}_\mem(t)\Big),
	\end{align}
	with the $n$-th derivative $F^{(n)} = \frac{\partial^n F(D_\theta)}{\partial D_\theta^n} \bigg|_{D_\theta=D_{\theta*}}.$ This allows us to write \begin{equation}
	\mu_\mem(D_\theta,\bm{x},t) = \tilde{\mu}^{(0)}_\mem(\bm{x},t) + {\mu}^{(1)}_\mem(\bm{x},t) D_\theta + \mathcal{O}((D_\theta-D_{\theta*})^2),
	\end{equation}
	with $\tilde{\mu}^{(0)}_\mem(\bm{x},t) = {\mu}^{(0)}_\mem(\bm{x},t) - {\mu}^{(1)}_\mem(\bm{x},t) D_{\theta*}.  $ \gj{Similarly, we also expand the average velocity $\bar{V}_x(D_\theta) = \tilde{V}_* + V_*^\prime D_\theta $. } Consistent with the main manuscript we have defined $\tilde{V}_*=\bar{V}_x(D_{\theta*}) - V_*^\prime D_{\theta*} $ and $V_*^\prime={\partial \bar{V}_x(D_\theta)}/{\partial D_\theta}|_{D_\theta=D_{\theta*}}$. Finally, we assume that $\varphi_0(D_\theta,\bm{x},t)$ is Gaussian and expand it in terms of its moments, $\rho(\bm{x},t) = \int dD_\theta \varphi_0(D_\theta,\bm{x},t),$ the average policy $\mu(\bm{x},t) = \rho(\bm{x},t)^{-1} \int dD_\theta D_\theta \varphi_0(D_\theta,\bm{x},t), $ and the diversity $\sigma^2(\bm{x},t) = \rho(\bm{x},t)^{-1}\int dD_\theta (D_\theta - \mu(\bm{x},t)) ^2 \varphi_0(D_\theta,\bm{x},t) $. This yields the analytical hydrodynamic equations,
	\begin{align}
	\frac{\partial \rho(\bm{x},t)}{\partial t} &= - \nabla  \gj{\cdot} \big( \gj{\tilde{\bm V}_* + {\bm V}_*^\prime} \mu(\bm{x},t) - D_0 \nabla  \big) \rho(\bm{x},t) ,  \\
	\frac{\partial \mu(\bm{x},t)}{\partial t} &= {- \big(\gj{\tilde{\bm V}_* + {\bm V}_*^\prime} \mu(\bm{x},t)\big) \gj{\cdot} \nabla \mu(\bm{x},t) -  \gj{{\bm V}_*^\prime \cdot} \nabla \sigma^2(\bm{x},t) -  \rho^{-1}(\bm{x},t) \sigma^2(\bm{x},t) \gj{{\bm V}_*^\prime \cdot} \nabla \rho(\bm{x},t) } + D_0 \Delta \mu(\bm{x},t) \nonumber \\
	& { + {2D_0}\rho^{-1}(\bm{x},t)  \nabla \mu(\bm{x},t) \gj{\cdot} \nabla \rho(\bm{x},t)}
	- 2\lamLt \mu^{(1)}_\mem(t) \sigma^2(\bm{x},t) \rho(\bm{x},t) \Big(  \tilde{\mu}^{(0)}_\mem(t) - V_T + \mu^{(1)}_\mem(t) \mu(\bm{x},t)  \Big) , \label{eq:hydromean}\\
	\frac{\partial \sigma^2(\bm{x},t)}{\partial t} &= {- \big(\gj{\tilde{\bm V}_* + {\bm V}_*^\prime} \mu(\bm{x},t)\big) \gj{\cdot} \nabla \sigma^2(\bm{x},t) - 2 \sigma^2(\bm{x},t) \gj{{\bm V}_*^\prime \cdot} \nabla \mu(\bm{x},t)} +  D_0 \Delta \sigma^2(\bm{x},t) \nonumber\\
	&{ + {2D_0}\rho^{-1}(\bm{x},t) \nabla \sigma^2(\bm{x},t) \gj{\cdot} \nabla \rho(\bm{x},t) + 2 D_0 |\nabla \mu(\bm{x},t)|^2} + 2 D_{\rm mut} - 2 \lamLt \rho(\bm{x},t) ({\mu^{(1)}_\mem(\bm{x},t)})^2  \sigma^4(\bm{x},t) \label{eq:hydrodiv} ,
	\end{align}
	\gj{where $\tilde{\bm V}_*=(\tilde{V}_*,0)$ and ${\bm V}_*^\prime=(V_*^\prime,0)$.}
	These equations combine the convection and diffusion induced transport of policies and diversity, as well as the teaching of new policies in the terms including $\lamLt$. Connecting the equations to the kinetic theory results shown in the main manuscript, Eqs.~(9) and (10), we therefore find,
	\begin{align}
	F_1(\mu,\sigma^2;\mu_{\mem}^{(n)}) &= - 2\lamLt \mu^{(1)}_\mem \Big(  \tilde{\mu}^{(0)}_\mem - V_T + \mu^{(1)}_\mem \mu  \Big)\\
	F_2(\mu,\sigma^2;\mu_{\mem}^{(n)}) &= 2 \lamLt  ({\mu^{(1)}_\mem})^2  .
	\end{align}
	
	\subsection{Analytical solutions for fixed targets}
	
	In the following we assume that both the target velocity $V_T$  and the initial conditions are independent of position $\bm{x}$. Furthermore, we assume timescale separation between $\tau_\mem$ and $\lambda_T$ and thus replace $\mu_\mem^{(n)}(t)$ by their steady-state values. We thus find,
	\begin{align}
	\frac{d \mu(t)}{d t} &= 
	- 2\lamLt {V_*^\prime}^2 \sigma^2(t) \rho \Big( \mu(t) - D_{\theta,T}   \Big) \\
	\frac{d \sigma^2(t)}{d t} &= 2 D_{\rm mut}  - 2 \lamLt \rho {V_*^\prime}^2 \sigma^4(t), 
	\end{align}
	where $D_{\theta,T} = D_{\theta*} + (V_T - \bar{V}_x(D_{\theta*}))/ V_*^\prime $ is an approximation of the target policy which maximizes the reward. In particular, if $\bar{V}_x(D_{\theta*}) \rightarrow V_T$ the relation becomes exact. Thus it is important to choose an appropriate $D_{\theta*}$.
	From these dynamical equations we can directly derive Eqs.~(12) and (13) as stated in the main manuscript.
	
	\subsection{Analytical solution for space-dependent targets}
	
	We are also interested in the steady-state solution in the presence of a space-dependent target $V_T(x) = V_T^s + V^\Delta_T \sin(2 \pi x / L)$. We assume that $D_0/L^2$ is significantly smaller than $v_0 /L$ and thus convective transport dominates. Furthermore, we expect that the spatial dependency of $\rho(\bm{x},t)$ and $\sigma^2(\bm{x},t)$ are significantly smaller than the one of $\mu(\bm{x},t)$. Finally, we assume that the convection velocity in the steady-state $\tilde{V}_* + V_*^\prime \mu(\bm{x},t) \approx V_T^s$, thus neglecting higher-order terms. Under these approximations we can derive that $\sigma^2(\bm{x},t) = \sqrt{\frac{2 D_{\rm mut}}{\lambda_0}}$ with $\lambda_0 = 2 \lamLt \rho {V_*^\prime}^2$ and thus find the partial differential equation,
	\begin{align}\label{eq:ansatz_space}
	0 = - V_T^s \frac{\partial \mu(x) }{\partial x} - \sqrt{2 D_{\rm mut} \lambda_0} (\mu(x) - D_{\theta,T}(x)),
	\end{align}
	where $D_{\theta,T}(x) = D_{\theta*} + (V_T(x) - \bar{V}_x(D_{\theta*}))/ V_*^\prime $. This equation corresponds to Eq.~(14) in the main manuscript. 
	
	To solve this equation we are using the ansatz, $\mu(x) = \mu_T^s + \mu^\Delta_T \sin(2 \pi x / L - \phi)$ \gj{and insert it into Eq.~(\ref{eq:ansatz_space}). Subsequently, we apply the trigonometric identities, $\sin(\alpha + \beta)=\sin(\alpha)\cos(\beta)+\cos(\alpha)\sin(\beta)$ and $\cos(\alpha + \beta)=\cos(\alpha)\cos(\beta)-\sin(\alpha)\sin(\beta)$ to separate the phase $\phi$ from the space-dependent terms. The resulting equation features orthogonal terms which are either constant (yielding $\mu_T^s$), proportional to $\cos(2 \pi x / L)$ (yielding $\phi$) and proportional to $\sin(2 \pi x / L)$ (yielding $\mu^\Delta_T$). Therefore, we find,
		\begin{align}
		\mu_T^s &=  D_{\theta*} + \frac{V_T^s-\bar{V}_x(D_{\theta*})}{V_*^\prime}\\
		\tan(\phi) &= \frac{2 \pi V_{T}^s}{ \sqrt{2 D_{\rm mut} \lambda_0} L}\\
		\mu^\Delta_{T} &= \frac{V^\Delta_T}{V_*^\prime} \cos(\phi).
		\end{align}}
	Finally, we identify the average velocity using the same approximations as above for constant $V_T$, $\langle V_x(x) \rangle = \bar{V}_x(D_{\theta*}) + V_*^\prime(\mu(x) - D_{\theta*})$, by inserting the above solution for $\mu(x),$ which finally yields the relations stated in the main manuscript.

	\subsection{Numerical details}
	
	\noindent To enable reproducibility we list in the following all the parameters used in the agent-based simulations, which are described in the main manuscript,
	\begin{itemize}
		\item $\tau_B = 0.004$
		\item $\sigma_B = 0.1$
		\item $\tau_\mem = 1.0$
		\item $\lambda_T^{-1} = 100$
		\item $v_0=1$, $D_0=0.0001$
		\item $\alpha_T=10$, corresponding to a very steep activation function. 
		\item $N=10000$
		\item $L_x=L_y=10 \rightarrow \rho=100$. Only for Fig.~3 we have set $L_x=100$ keeping the density constant, i.e. choosing $N=10^5$.
		\item $\Delta t = 0.002$ which is small enough to capture the smallest timescales.
		\item $r_{c} = \frac{1}{\sqrt{\pi}}$, to ensure that each particle has approximately $ \rho$ particles in its communication neighborhood. This ensures consistency with the kinetic theory. 
		\item $\mu(t=0) = 100$, $\sigma^2(t=0) = 400.$
		
	\end{itemize}
	
	\section{Derivations for MODEL2: Robots}
	\label{sec:mod2}
	
	\begin{figure}
		\includegraphics{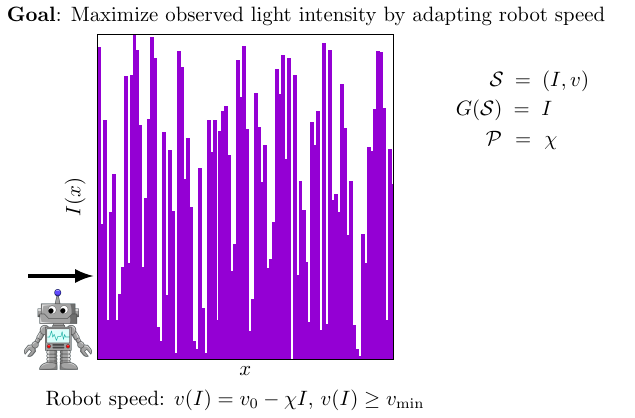}
		\caption{Illustration of MODEL2, featuring non-interacting robots which attempt to maximize the light-intensity $\bar{I}(\chi)$ to which they are exposed by teaching the optimal parameter $\chi$ controlling their local movement speed. }
	\end{figure}
	
	The second model represents a robot moving in a landscape of space-dependent external light intensity. The model is one-dimensional and we assume that space is separated into $N_{\rm bin}$  equally-sized bins $i$ of size $\Delta L$. Each bin has light intensity $I_i$, which are uncorrelated in space and equally distributed in $I_i \in [0,I_{\rm max}]$ thus forming the intensity field $I(x),$ which can be measured by the robot. The speed of the robot is then defined as $v(I)= v_0 - \chi I$ and capped at $v_\text{min} > 0$ (thus the robot always moves forward). The parameter $\chi$ thus defines the sensitivity with which the robot reacts to the external light. The state of the robot is therefore $\mathcal{S}=(I,v)$. We apply periodic boundary conditions, but generally assume $L$ to be large compared to $\Delta L$. We assume that we have $N$ robots in the world, but robots do not interact with each other (apart from teaching). The goal of each robot is to adjust the policy $\policy=\chi$ such that it maximizes $\bar{I}(\chi)$, i.e., the average light intensity seen by a robot with sensitivity $\chi$.
	
	\subsection{Solving the microscopic physical model}
	
	The probability for each robot to be in bin $i$ with intensity $I_i$ is,
	\begin{equation}
	p(I) \propto v^{-1}(I) = \left\{\begin{array}{lr}
	(v_0 - \chi I)^{-1}, & \text{for } I < I_*\\
	v_\text{min}^{-1}, & \text{for }I \geq I_*
	\end{array}\right.
	\end{equation}
	with $I_*=\frac{v_0 - v_\text{min}}{\chi}$. Assuming that the physical time scale $\Delta L / v_0$ is much smaller than any other time scale in the system, i.e., that each robot sees many different $I_i$ on the memory time scale $\tau_\mem$, we can then calculate the expected average light-intensity seen by the robot with policy $\chi$,
	\begin{equation}
	\bar{I}(\chi) = \frac{ \int_0^{I_\text{max}} d I\, I\, p(I) }{\int_0^{I_\text{max}} d I\, p(I) }.
	\end{equation}
	The integral can be separated into two parts going from $[0,I_*]$ and $[I_*, {I}_{\rm max}]$ and then solved easily. The final result is,
	{\small 
		\begin{equation}
		\bar{I}(\chi) = \left\{\begin{array}{ll}
		\left[\frac{1}{2 v_\text{min}} \left( I_\text{max}^2 - {I_*}^2 \right) -\frac{I_*}{\chi} + \frac{v_0}{\chi^2} \ln \left( v_0 /v_\text{min} \right)   \right]\left[ \frac{I_\text{max} - I_*}{v_\text{min}} + \frac{1}{\chi} \ln \left(\frac{v_0}{v_{\rm min}}\right) \right]^{-1} & , \text{ for } \chi > \frac{v_0 - v_\text{min}}{I_\text{max}}\\
		\frac{v_0}{\chi} - \frac{I_\text{max}}{\ln(v_0/(v_0-\chi I_\text{max}))}  & , \text{ otherwise}
		\end{array}\right.
		\end{equation}
	}
	In MODEL2 we choose $G(\mathcal{S})=I$, thus $\bar{I}(\chi)$ directly corresponds to the term $\bar{G}(\policy)$ introduced in the main manuscript. Different from MODEL1 we then choose as reward function $\reward(\mem) =\mem$ since we aim at maximizing the collected light intensity. We thus find for Eqs.~(23) and (24) in the main manuscript (dropping any space dependence on longer time scales and assuming time scale separation),
	\begin{align}
	\label{eq:MODEL2rho}
	\frac{\partial \varphi_0(\chi,t)}{\partial t} &=  \lamLt \varphi_0 \rho(t)   \mu_\mem \gj{-} \lamLt \varphi_0 \int d\chi^\prime\varphi_0(\chi^\prime,t)  \mu_\mem(\chi^\prime,t)    + D_{\rm mut} \frac{\partial^2}{\partial \chi^2} \varphi_0  \\
	\frac{\partial \mu_\mem(\chi,t)}{\partial t}&= \tau_\mem^{-1} \Big(\bar{I}(\chi) - \mu_\mem\Big).
	\end{align}
	To capture the details of $\bar{I}(\chi)$ we expand it to fourth order. As before, we expand around a predefined policy $\chi_*,$
	\begin{align}
	\frac{\partial \mu^{(n)}_\mem(t)}{\partial t}&= \tau_\mem^{-1} \Big(\bar{I}^{(n)} - \mu^{(n)}_\mem(t)\Big),
	\end{align}
	with the $n$-th derivative $F^{(n)} = \frac{\partial^n F(\chi)}{\partial \chi^n} \bigg|_{\chi=\chi_*}.$ This allows us to write
	\begin{align}
	\mu_\mem(\chi,t) &= \tilde{\mu}^{(0)}_\mem(t) + \tilde{\mu}^{(1)}_\mem(t) \chi - \tilde{\mu}^{(2)}_\mem(t) \chi^2 +  \tilde{\mu}^{(3)}_\mem(t) \chi^3 - \tilde{\mu}^{(4)}_\mem(t) \chi^4 , \text{ with}\\
	\tilde{\mu}^{(0)}_\mem(t) &= {\mu}^{(0)}_\mem(t) - {\mu}^{(1)}_\mem(t) \chi_* + \frac{1}{2}{\mu}^{(2)}_\mem(t) \chi_*^2 - \frac{1}{6}{\mu}^{(3)}_\mem(t) \chi_*^3 + \frac{1}{24} {\mu}^{(4)}_\mem(t) \chi_*^4,\\
	\tilde{\mu}^{(1)}_\mem(t) &=  {\mu}^{(1)}_\mem(t) -  {\mu}^{(2)}_\mem(t) \chi_* + \frac{1}{2}{\mu}^{(3)}_\mem(t) \chi_*^2 - \frac{1}{6} {\mu}^{(4)}_\mem(t) \chi_*^3\\
	\tilde{\mu}^{(2)}_\mem(t) &= - \frac{1}{2}{\mu}^{(2)}_\mem(t) + \frac{1}{2}{\mu}^{(3)}_\mem(t) \chi_* - \frac{1}{4} {\mu}^{(4)}_\mem(t) \chi_*^2,\\
	\tilde{\mu}^{(3)}_\mem(t) &=   \frac{1}{6}{\mu}^{(3)}_\mem(t)  - \frac{1}{6} {\mu}^{(4)}_\mem(t) \chi_*\\
	\tilde{\mu}^{(4)}_\mem(t) &=  - \frac{1}{24} {\mu}^{(4)}_\mem(t) .
	\end{align}
	Inserting this expansion into Eq.~(\ref{eq:MODEL2rho}) and using the same moments, average policy $\mu$ and diversity $\sigma^2$, as for MODEL1, we finally find,
	\begin{align}
	\frac{\partial \mu(t)}{\partial t} &= 
	\lamLt \rho \sigma^2(t)  \Big(  \tilde{\mu}^{(1)}_\mem(t) - 2 \tilde{\mu}^{(2)}_\mem(t) \mu(t)  \gj{+3\tilde{\mu}^{(3)}_\mem(t)(\mu^2(t)+\sigma^2(t))-4\tilde{\mu}^{(4)}_\mem(t) \mu(t)(\mu^2(t)+3\sigma^2(t))}  \Big) \\
	\frac{\partial \sigma^2(t)}{\partial t} &= 2 D_{\rm mut}  - 2 \lamLt \rho \sigma^4(t) \Big( {\tilde{\mu}^{(2)}_\mem(t)} - 3 \tilde{\mu}^{(3)}_\mem(t) \mu(t) + 6  \tilde{\mu}^{(4)}_\mem(t) \big(\mu^2(t) +  \sigma^2(t) \big)  \Big).
	\end{align}
	These equations have been used for the numerical results shown in Fig.~4 in the main manuscript. 
	
	\subsection{Numerical details}
	
	\noindent The following parameters were used in the agent-based simulations,
	\begin{itemize}
		\item $\Delta L = 10^{-3}$
		\item $\tau_\mem = 1.0$
		\item $\lambda_T^{-1} = 100$
		\item $v_0=1$, $v_{\rm min}=0.01$
		\item $D_0=0$
		\item $\alpha_T=10$, corresponding to a very steep activation function. 
		\item $N=10000$
		\item $L=100 \rightarrow \rho=100$. 
		\item $\Delta t = 0.0001$ which is small enough to capture the smallest timescales.
		\item $r_{c} = 0.5$, to ensure that each particle has approximately $ \rho$ particles in its communication neighborhood. This ensures consistency with the kinetic theory.
		\item $\mu(t=0) = 1.25$, $\sigma^2(t=0) = 0.0225.$
		
	\end{itemize}
	
	\section{Multi-dimensional Kinetic Theory}
	\label{sec:multidim}
	
	We have emphasized in the main manuscript that the state $\state$, the memory $\mem$ and the policy $\policy$ could be multi-dimensional. Here, we show that one can similarly derive hydrodynamic equation in such a situation. (Ignoring space to keep the notation more compact.) \gj{Each component in $\state$, $\mem$, and $\policy$, will be specified by the indices $i$, $j$, $k$, $l$, $n$, etc.}
	
	The multi-dimensionality starts to play a role once we introduce the moments in the memory, including the density,  $\varphi_0(\policy,t) = \int d \mem  f_0(\mem,\policy,t) $,  the average memory, $\mu_{\mem,i}(\policy,t) = \varphi_0^{-1} \int d \mem \mem_i f_0(\mem,\policy,t) $ and the memory covariance matrix, $\sigma^2_{\mem,ij}(\policy,t) = \varphi_0^{-1} \int d \mem (\mem_i - \mu_{\mem,i}(\policy,t) )(\mem_j - \mu_{\mem,j}(\policy,t) ) f_0(\mem,\policy,t) $. As in MODEL1 we assume that $\reward(\mem)= - \sum_i (\mem_i - T_i)^2$, i.e., we want that each entry in the memory approaches a certain target $T_i.$ Assuming the time scale hierarchy, $\tau_G \ll \tau_\mem \ll \tau_T$, we then find for the policy dynamics,
	\begin{align}
	\frac{\partial \varphi_0(\policy,t)}{\partial t} &=\lamLt \varphi_0 \int d\policy^\prime\varphi_0(\policy^\prime,t)\sum_i  \Big(  \sigma^2_{\mem,ii}(\policy^\prime,t)+ \mu_{\mem,i}(\policy^\prime,t)^2 -2 T_i \mu_{\mem,i}(\policy^\prime,t) \Big) \nonumber \\
	&+\lamLt \varphi_0 \rho(t) \sum_i \Big(   2 T_i  \mu_{\mem,i} -\mu_{\mem,i}^2 - \sigma^2_{\mem,ii} \Big) + D_{\rm mut} \gj{\sum_i \frac{\partial^2}{\partial \policy_i^2}} \varphi_0  \\
	\frac{\partial \mu_{\mem,i}(\policy,t)}{\partial t}&=  \tau_\mem^{-1} \Big(\bar{G}_i(\policy) - \mu_{\mem,i}\Big) .
	\end{align}
	As in the main manuscript, we denote as $\bar{G}_i(\policy)=\langle G_i(\state) \rangle_\policy $ the policy-dependent ensemble average of the information $ G_i(\state). $ \gj{The specific choice of the reward $\reward(\mem)$ implies that there are no direct interactions between different components of the memory, i.e. the off-diagonal components $\sigma^2_{\mem,ij}$ are irrelevant. Including additional mixed terms in the reward, would therefore induce additional interactions.}
	
	We can then expand the memory $\mu_{\mem,i}(\mathcal{P},t)$ around a predefined policy $D_{\theta*}=\policy_*,$ by introducing the quantities $\mu_{\mem,i}^{(0)} =  \mu_{\mem,i}(\policy_{*})$, $\mu^{(1)}_{\mem,ij}(t)= \frac{\partial  \mu_{\mem,i}(\policy) }{\partial \policy_j} \bigg|_{\policy=\policy_*},$ $\bar{G}_{*i} =  \bar{G}_i (\policy_{*})$ and $\bar{G}^\prime_{*,ij}= \frac{\partial  \bar{G}_i(\policy) }{\partial \policy_j} \bigg|_{\policy=\policy_*}.$ We thus find,
	\begin{align}
	\frac{\partial \mu^{(0)}_{\mem,i}(t)}{\partial t}&= \tau_\mem^{-1} \Big(\bar{G}_{*i} - \mu^{(0)}_{\mem,i}(t)\Big),\\
	\frac{\partial \mu^{(1)}_{\mem,ij}(t)}{\partial t}&= \tau_\mem^{-1} \Big(\bar{G}^\prime_{*ij} - \mu^{(1)}_{\mem,ij}(t)\Big).
	\end{align}
	This allows us to expand the memory, $\mu_{\mem,i}(\policy,t) = \tilde{\mu}^{(0)}_{\mem,i}(t) + {\mu}^{(1)}_{\mem,ij}(t) \policy_j,$ with $\tilde{\mu}^{(0)}_{\mem,i}(t) = {\mu}^{(0)}_{\mem,i}(t) - {\mu}^{(1)}_{\mem,ij}(t) D_{\theta*,j},  $ using Einstein summation convention. 
	
	As described around Eq.~(\ref{eq:mem2}), we find that $\sigma_\mem^2$ is independent of $\policy$ when assuming timescale separation and thus can be canceled out. Finally, we assume that $\varphi_0(\policy,t)$ is Gaussian and expand it in terms of its moments, the average policy $\mu_i(t) = \rho(t)^{-1} \int d\policy \policy_i \varphi_0(\policy,t), $ and the covariance $\sigma_{ij}^2(t) = \rho(t)^{-1}\int d\policy (\policy_i - \mu_i(t)) (\policy_j - \mu_j(t)) \varphi_0(\policy,t) $. We also identify higher moments for multi-dimensional Gaussians
	\begin{align}
	& \int d\policy (\policy_i - \mu_i(t)) (\policy_j - \mu_j(t)) (\policy_k - \mu_k(t))\varphi_0(\policy,t) = 0 \\
	\Rightarrow &\rho(t)^{-1} \int d\policy \policy_i \policy_j \policy_k \varphi_0(\policy,t)  = \mu_i(t)\mu_j(t)\mu_k(t) + \mu_i(t)\sigma^2_{jk}(t) + \mu_j(t)\sigma^2_{ik}(t) + \mu_k(t)\sigma^2_{ij}(t) \\
	& \int d\policy (\policy_i - \mu_i(t)) (\policy_j - \mu_j(t)) (\policy_k - \mu_k(t)) (\policy_l - \mu_l(t))\varphi_0(\policy,t) = \sigma^2_{ij}(t)\sigma^2_{kl}(t) + \sigma^2_{ik}(t)\sigma^2_{jl}(t) + \sigma^2_{il}(t)\sigma^2_{kj}(t) \\
	\Rightarrow &\rho(t)^{-1} \int d\policy \policy_i \policy_j \policy_k \policy_l \varphi_0(\policy,t)  = \mu_i(t)\mu_j(t)\mu_k(t)\mu_l(t) + \sigma^2_{ij}(t)\sigma^2_{kl}(t) + \sigma^2_{ik}(t)\sigma^2_{jl}(t) + \sigma^2_{il}(t)\sigma^2_{kj}(t) + \mu_i(t)\mu_j(t)\sigma^2_{kl}(t) \nonumber\\
	&+ \mu_i(t)\mu_k(t)\sigma^2_{jl}(t) 
	+ \mu_i(t)\mu_l(t)\sigma^2_{jk}(t) + \mu_j(t)\mu_k(t)\sigma^2_{il}(t)+ \mu_j(t)\mu_l(t)\sigma^2_{ik}(t)+ \mu_k(t)\mu_l(t)\sigma^2_{ij}(t).
	\end{align} 
	After some calculations we find,
	\begin{align}
	\frac{\partial \mu_l(t) }{\partial t} &= -2\lamLt \rho  \mu_{\mem,i j}^{(1)} \sigma^2_{jl}(t) \left[ \mu_{\mem,i k}^{(1)}  \mu_k(t)   -  T_i  +  \tilde{\mu}_{\mem,i}^{(0)}  \right] , \\
	\frac{\partial \sigma^2_{ln}(t) }{\partial t} &= 2 D_{\rm mut} \delta_{ln} -2\lamLt \rho  \mu_{\mem,i j}^{(1)} \mu_{\mem,i k}^{(1)}  \sigma^2_{lj}\sigma^2_{nk} .
	\end{align}
	While these equations are slightly more complex than the one-dimensional Eqs.~(\ref{eq:hydromean}) and (\ref{eq:hydrodiv}), they still share very similar characteristics. For example, if the covariance for dimension $l$ of the $\policy_l$ is zero, i.e., $\sigma^2_{kl}=0\, \forall k,$ we find immediately that ${\partial \mu_l(t) }/{\partial t}=0$ as expected from Fisher's first theorem of natural selection \cite{ewens1989interpretation}. \gj{ Despite the apparent independence of the different components of the memory, as discussed above, they can indirectly influence each other. For example, one component could influence the reward and thus an emergent policy. This policy will then influence, via the processed information $G_i(\state),$ other components of the memory. The long-term policy that emerges can thus include very non-trivially coupling between different components of the memory.  }
	
	\bibliography{library_local.bib}